\documentclass{aa}

\usepackage{graphicx}
\usepackage{txfonts}
\begin{document}

\title{Sunspot penumbral filaments intruding into a light bridge and the resultant reconnection jets}

\author{Y. J. Hou \inst{1,2}
      \and T. Li \inst{1,2}
      \and S. H. Zhong \inst{1,2}
      \and S. H. Yang \inst{1,2}
      \and Y. L. Guo \inst{1,2}
      \and X. H. Li \inst{1,2}
      \and J. Zhang \inst{3,1}
      \and Y. Y. Xiang \inst{4}
      }

\institute{CAS Key Laboratory of Solar Activity, National Astronomical Observatories,
           Chinese Academy of Sciences, Beijing 100101, China; yijunhou@nao.cas.cn\\
           \and School of Astronomy and Space Science, University of Chinese Academy of Sciences, Beijing 100049, China\\
           \and School of Physics and Materials Science, Anhui University, Hefei 230601, China; zjun@ahu.edu.cn\\
           \and Yunnan Observatories, Chinese Academy of Sciences, Kunming 650011, China
           }

\date{Received 16 June 2020 / Accepted 29 July 2020}

% \abstract{}{}{}{}{}
% 5 {} token are mandatory

  \abstract
  % context heading (optional)
  % {} leave it empty if necessary
   {Penumbral filaments and light bridges are prominent structures inside sunspots and are important for understanding
   the nature of sunspot magnetic fields and magneto-convection underneath.
   }
  % aims heading (mandatory)
   {We investigate an interesting event where several penumbral filaments intruded into a sunspot light bridge for more
   insights into magnetic fields of the sunspot penumbral filament and light bridge, as well as their interaction.
   }
  % methods heading (mandatory)
   {Combining data from the New Vacuum Solar Telescope, \emph{Solar Dynamics Observatory}, and \emph{Interface Region
   Imaging Spectrograph}, we study the emission, kinematic, and magnetic topology characteristics of the penumbral
   filaments intruding into the light bridge and the resultant jets.
   }
  % results heading (mandatory)
   {At the west part of the light bridge, the intruding penumbral filaments penetrated into the umbrae on both sides of
   the light bridge, and two groups of jets were also detected. The jets shared the same projected morphology with the
   intruding filaments and were accompanied by intermittent footpoint brightenings. Simultaneous spectral imaging
   observations provide convincing evidences for the presences of magnetic reconnection related heating and bidirectional
   flows near the jet bases and contribute to measuring vector velocities of the jets. Additionally, nonlinear force-free
   field extrapolation results reveal strong and highly inclined magnetic fields along the intruding penumbral filaments,
   consistent well with the results deduced from the vector velocities of the jets. Therefore, we propose that the jets
   could be caused by magnetic reconnections between emerging fields within the light bridge and the nearly horizontal
   fields of intruding filaments. They were then ejected outward along the stronger filaments fields.
   }
  % conclusions heading (optional), leave it empty if necessary
   {Our study indicates that magnetic reconnection could occur between the penumbral filament fields and emerging fields
   within light bridge and produce jets along the stronger filament fields. These results further complement
   the study of magnetic reconnection and dynamic activities within the sunspot.
   }


\titlerunning{sunspot penumbral filaments, light bridges, and jets}
\authorrunning{Hou et al.}

\maketitle
%
%________________________________________________________________

\section{Introduction}
Sunspots are manifestations of high concentrations of solar magnetic field, where the strong magnetic field
inhibits the normal convective transports of plasma and energy from the solar interior to the surface (Gough \& Tayler
1966; Solanki 2003; Borrero \& Ichimoto 2011). A typical sunspot is composed of a dark umbra and a filamentary penumbra
surrounding the umbra. The sunspot magnetic fields are believed to be near vertical in the umbra and more inclined in
the penumbra (Thomas et al. 2002). Inside the dark sunspots, many small-scale structures with inhomogeneous brightness
have been studied, such as umbral dots (Beckers \& Schr{\"o}ter 1968; Sobotka et al. 1997; Riethm{\"u}ller et al. 2008;
Goodarzi et al. 2018), light bridges (Muller 1979; Lites et al. 1991; Sobotka et al. 1994; Leka 1997; Berger \& Berdyugina
2003; Katsukawa et al. 2007b; Rouppe van der Voort et al. 2010; Lagg et al. 2014; Felipe et al. 2017; Wang et al. 2018;
Dur{\'a}n et al. 2020), penumbral filaments (Muller 1973; Schlichenmaier et al. 1998; Scharmer et al. 2002; Ichimoto
et al. 2007; Langhans et al. 2007; Su et al. 2010; Tiwari et al. 2013), and umbral filaments (Kleint \& Sainz Dalda 2013;
Guglielmino et al. 2017, 2019). Umbral dots are bright dot-like features embedded in the umbral background, and their
number rapidly increases with decreasing diameter (Sobotka et al. 1997). Light bridges are bright lanes penetrating into
the dark umbra, and strong ones can even completely separate the umbra. Previous studies showed that the magnetic fields
of light bridges are generally weaker and more inclined than the neighboring strong and vertical umbral fields, forming
a magnetic canopy (Rueedi et al. 1995; Jur{\v c}{\'a}k et al. 2006; Felipe et al. 2016). Recent observations and
simulations have revealed the existence of weakly twisted and emerging magnetic fields in the light bridge (Louis et al.
2015; Toriumi et al. 2015a, b; Yuan \& Walsh 2016). Penumbral filaments are radial structures in the penumbra,
consisting of a dark core flanked by lateral brightenings (Scharmer et al. 2008). Spruit \& Scharmer (2006)
introduced the concept of field-free convection gaps, which intrude into the strong sunspot magnetic field above and
form a cusp-like structure near the observed surface ($\tau=1$), to explain the formations of umbral dots, light
bridges, and penumbral filaments. And a strong horizontal magnetic field is predicted to exist along the penumbral
filament. As for umbral filaments, they are unusual elongated filamentary bright structures with strong horizontal
fields within sunspot umbrae, and are interpreted as the photospheric manifestation of a flux rope hanging above the
umbra (Guglielmino et al. 2019).

The structures mentioned above are usually derived from local inhomogeneous magnetic field and magneto-convection,
which indicate the complexity of sunspot magnetic and thermal properties (Spruit \& Scharmer 2006; Sch{\"u}ssler
\& V{\"o}gler 2006; Rimmele 2008; Rempel et al. 2009; Reardon et al. 2013). As a result of the deviation from sunspot
background conditions, various dynamic phenomena especially surge-like activities frequently occur around these
structures, such as surges and light walls above the light bridges (Roy 1973; Asai et al. 2001; Shimizu et al. 2009;
Louis et al. 2014a; Yang et al. 2015, 2016, 2017; Bharti 2015; Robustini et al. 2016; Hou et al. 2016b, 2017; Zhang
et al. 2017; Tian et al. 2018; Bai et al. 2019), transient jets above a penumbral filament intrusion into the umbra
(Bharti et al. 2017), and ubiquitous penumbral microjets in the penumbral chromosphere (Katsukawa et al. 2007a;
Jur{\v{c}}{\'a}k \& Katsukawa 2008; Nakamura et al. 2012; Tiwari et al. 2016; Drews \& Rouppe van der Voort 2017;
Samanta et al. 2017; Esteban Pozuelo et al. 2019; Rouppe van der Voort \& Drews 2019). These surge-like activities
display an impressive variety of morphological, temporal, and spectral properties.

Some of these surge-like activities are oscillatory in nature, such as the bright wall-shaped structures: light
walls, which are recently revealed by high-resolution observations from the \emph{Interface Region Imaging
Spectrograph} (\emph{IRIS}; De Pontieu et al. 2014). The most prominent feature of light walls is their coherent
oscillating bright fronts in 1330/1400 {\AA} channel, which have rising speeds of about 10--20 km s$^{-1}$ (Yang
et al. 2015). Zhang et al. (2017) found that the core of the \emph{IRIS} Mg {\sc ii} k 2796.35 {\AA} line within
an oscillating light wall above a light bridge repeatedly experiences a fast, impulsive blueward excursion followed
by gradual motion to a redshift. Then the authors proposed that these oscillating walls result from upward shocked
p-mode waves. The similar viewpoint is also discussed in works about sunspots (Morton 2012; Rouppe van der Voort
\& de la Cruz Rodr{\'{\i}}guez 2013; Hou et al. 2017; Tian et al. 2018; Hou et al. 2018). Through investigating two
special events where the light walls co-existed with traditional surges (or jets) above light bridges, Hou et al.
(2017) compared the two different phenomena in aspects of rising velocities, heights, lifetimes, and base widths
and confirmed p-mode shock waves again as the source of the oscillating light walls. They noticed that the
oscillating light walls rise along the vertical umbral field to a typical height of several Mm with a projected
velocity of about 10 km s$^{-1}$, and could exist above the light bridge over a spatial range of about tens of Mm
for several hours (even up to several days in Hou et al. 2016b).

On the other hand, a considerable part of these surge-like activities within sunspots displays obvious features
of magnetic reconnection, such as the intermittent surges above sunspot light bridges with high speeds ($\sim$100
km s$^{-1}$). These surges have been intensively investigated through observations in H$\alpha$ and Ca {\sc ii}
channels and believed to be driven by magnetic reconnection between emerging magnetic arcades within the light
bridge and surrounding vertical umbral fields (Asai et al. 2001; Shimizu et al. 2009; Louis et al. 2014a;
Robustini et al. 2016; Tian et al. 2018; Yang et al. 2019b). Using high-resolution imaging spectroscopy in H$\alpha$
from the Swedish 1 m Solar Telescope, Robustini et al. (2016) reported on fan-shaped jets (with lengths of 7--38
Mm and speed of $\sim$100 km s$^{-1}$) above a light bridge and interpreted magnetic reconnection as the driver of
these jets. Recently, at light bridges, Tian et al. (2018) detected frequently occurring fine-scale jets with an
inverted Y-shape and transient footpoint brightenings, which are indicative of magnetic reconnection. The
simultaneous \emph{IRIS} spectral line profiles measured at the jet footpoint brightenings show significant
broadenings and enhancements at both wings, strongly indicating the occurrence of magnetic reconnection. Similarly,
Bharti et al. (2017) reported a series of transient jets extending $\sim$3 Mm above a penumbral intrusion into a
sunspot umbra. The observed transient $\lambda$-shaped bases of these jets indicate that magnetic reconnection
between emerging magnetic arcades and the pre-existing more vertical umbral magnetic fields produces these jets,
which then are ejected along the vertical umbral fields. In addition, penumbral microjets, the short-lived
($\leq$1 minute) fine-scale (0.4 Mm in width and 1--4 Mm in length) jetlike features (apparent speed of $\geq$100
km s$^{-1}$) in chromospheres of sunspot penumbrae, are another kind of activity believed to be driven by magnetic
reconnection (Katsukawa et al. 2007a; Sakai \& Smith 2008; Magara 2010; Drews \& Rouppe van der Voort 2017).
Magnetic reconnection is believed to occur between the horizontal penumbral filament fields and more vertical
penumbral fields, and leads to the penumbral microjets, which are aligned with the relatively vertical penumbral
magnetic fields (Jur{\v{c}}{\'a}k \& Katsukawa 2008; Katsukawa \& Jur{\v{c}}{\'a}k 2010). Alternatively, Ryutova
et al. (2008) suggested that shocks generated by magnetic reconnection between neighboring penumbral filaments
produce penumbral microjets, which is supported by observations reported in Reardon et al. (2013).

As mentioned above, the surge-like activities within sunspots with features of reconnection are suggested to be
driven by the magnetic reconnection occurring either between emerging magnetic fields within the light bridge and
vertical background sunspot (umbral) fields (Shimizu et al. 2009; Robustini et al. 2016; Hou et al. 2017; Tian
et al. 2018) or between the horizontal penumbral filament fields and relatively vertical background sunspot
(penumbral) fields (Katsukawa et al. 2007a; Jur{\v{c}}{\'a}k \& Katsukawa 2008; Magara 2010). Here naturally come
two questions, will magnetic reconnection occur between the emerging fields within light bridge and the penumbral
filament fields? If yes, what would then happen? It seems self-evident that as long as the magnetic
topology and plasma conditions are conducive for reconnection through flux emergence, the reconnection would occur,
and the accelerated plasma from the reconnection site has to follow the field lines. Direct observations
are necessary but absent yet to answer the two questions. In the present work, based on high-quality observations
from the New Vacuum Solar Telescope (NVST; Liu et al. 2014) in China, \emph{IRIS}, and \emph{Solar Dynamics
Observatory} (\emph{SDO}; Pesnell et al. 2012), we investigate an interesting event that would contribute to
providing answers to the two questions. In this event, sunspot penumbral filaments with nearly horizontal strong
magnetic fields intrude into a sunspot light bridge. Then transient brightenings and jets are observed around
the intrusion site. Unlike the surge-like activities aligned with the sunspot vertical fields reported
previously, the intermittent jets emanating from the brightenings reported here are along the highly inclined
fields of intruding penumbral filaments. Although the spectral profiles detected at the jet footpoint brightenings
show similar features to those reported earlier, the brightenings and jets in our observations are believed to be
produced in a different magnetic configuration: through reconnection between the emerging fields within the light
bridge and the nearly horizontal fields of intruding penumbral filaments.

In the present work, combining successive imaging observations with dominant emissions sampled from different
heights and abundant spectral data obtained at specific sites, we investigate the temporal evolution and spectral
properties of sunspot penumbral filaments intruding into the light bridge and the resultant jets. Furthermore,
the observed photospheric vector magnetic fields and extrapolated three-dimensional (3D) fields through nonlinear
force-free field (NLFFF) modeling enable us to better understand magnetic morphological properties of
these intruding penumbral filaments and jets. A cartoon model for the formation mechanism of these jets and its
implication on magnetic fields of the sunspot penumbral filaments and light bridge are also proposed. The remainder
of this paper is organized as follows. Section 2 describes the observations and data analysis taken in our study.
In Sect. 3, we present the results of data analysis in detail. Finally, we summarize the major findings and discuss
the results in Sect. 4.

\section{Observations and data analysis}
\begin{figure*}
\centering
\includegraphics [width=0.76\textwidth]{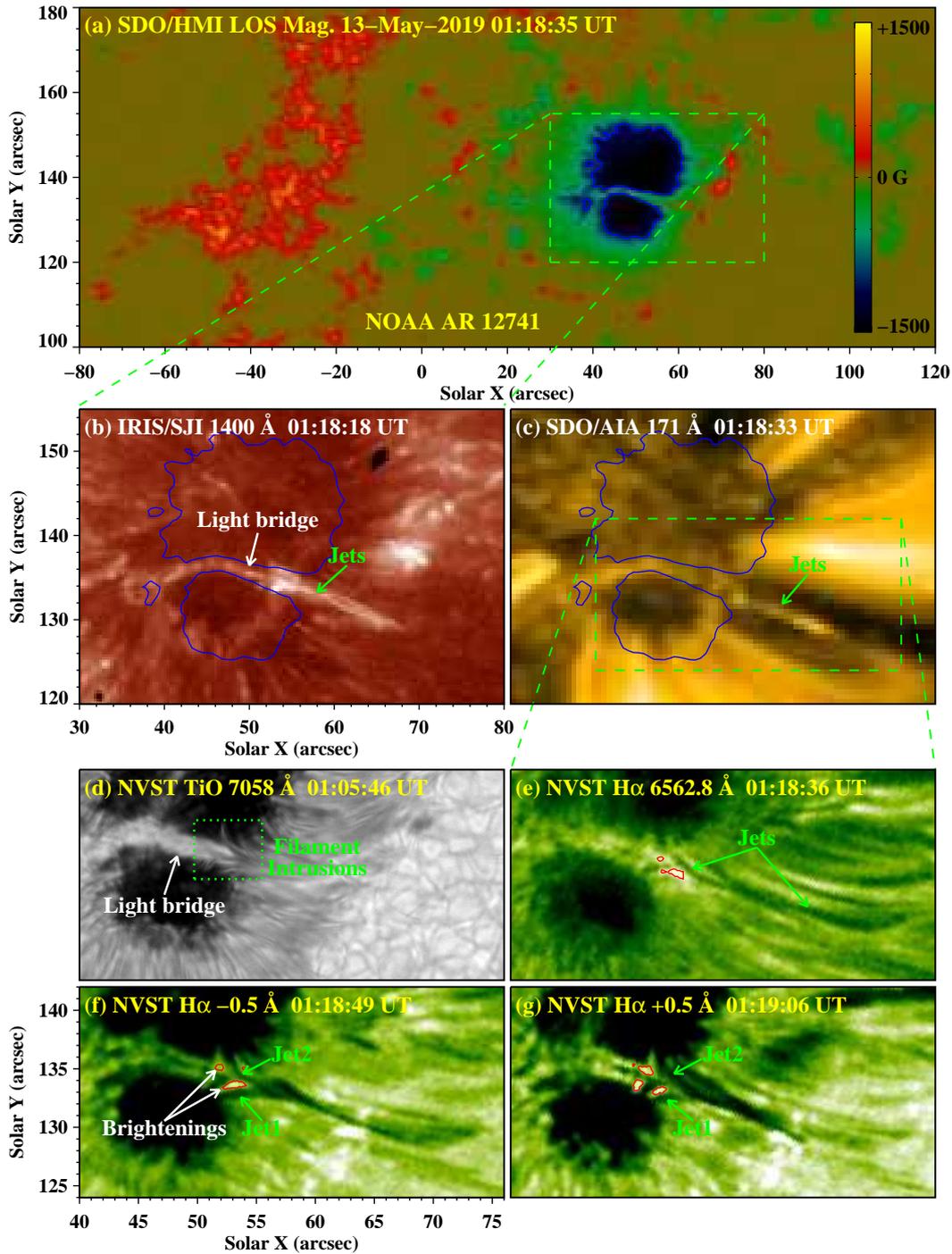}
\caption{Overview of the main sunspot, light bridge, penumbral filaments, and jets in AR 12741 on 2019 May 13.
(a)--(c): \emph{SDO}/HMI LOS magnetogram, \emph{IRIS}/SJI of 1400 {\AA}, and \emph{SDO}/AIA 171 {\AA} image showing
the magnetic fields of AR 12741 and the jets emanating from the west part of the light bridge within the main
sunspot. The blue curves in (a) are contours of the LOS magnetogram at --1000 G and are duplicated to (b) and (c).
The green dashed rectangle in (c) outlines the field of view (FOV) of (d)--(g) and Figs. 2--6.
(d): NVST TiO image displaying the penumbral filaments, which intrude into the light bridge and sunspot umbrae.
(e)--(g): Images of H$\alpha$ line core and two wings ($\Delta\lambda=\pm0.5$ {\AA}) exhibiting two groups of jets
(jet1 and jet2) emanating from the light bridge and brightenings near the jet bases (see the red contours).
An animation (figure1.mov) of 1400 {\AA} and 171 {\AA} images, covering 00:51 UT to 01:51 UT, is available in the
on-line journal.
}
\label{fig1}
\end{figure*}

Based on observations from the NVST, \emph{IRIS}, and \emph{SDO}, we investigate an event where the sunspot penumbral
filament intrusions into umbrae produce two groups of jets on both sides of a light bridge. This event occurred in
active region (AR) NOAA 12741 near the solar center on 2019 May 13. It was observed by the NVST from 00:54 UT to
03:05 UT in TiO 7058 {\AA} channel and H$\alpha$ channels (i.e., the line center at 6562.8 {\AA} and two line wings
at $\pm$ 0.5 {\AA}). The NVST is a vacuum solar telescope with a 985 mm clear aperture and is located in Fuxian Lake
of China, which has come into operation since 2012. The multi-channel high-resolution imaging system of NVST consists
of one channel for the chromosphere (H$\alpha$ 6562.8 {\AA}) and two channels for the photosphere (TiO 7058 {\AA} and
G-band 4300 {\AA}). The H$\alpha$ filter is a tunable Lyot filter with a bandwidth of 0.25 {\AA}. It can scan spectra
in the $\pm$ 5 {\AA} range with a step size of 0.1 {\AA} (Liu et al. 2014). The H$\alpha$ images employed here have a
field of view (FOV) of $\sim178{\arcsec}\times179{\arcsec}$, a pixel size of 0.{\arcsec}165, and a cadence of $\sim$43
s. The TiO images are obtained with a spatial sampling of 0.{\arcsec}052 pixel$^{-1}$, a cadence of 30 s, and a FOV of
$\sim143{\arcsec}\times118{\arcsec}$. The observed H$\alpha$ and TiO images are firstly calibrated through dark current
subtraction and flat field correction, and then reconstructed to Level 1+ by speckle masking (Xiang et al. 2016).

Moreover, the \emph{IRIS} was also focused on this region from 00:25 UT to 01:51 UT, and the spectral data were taken
in a very large dense 320-step raster mode with a step scale of 0.{\arcsec}35 and a step cadence of 16.2 s. Level 2
\emph{IRIS} data are employed here, which have been calibrated through dark current subtraction, flat field, geometrical,
and orbital variation corrections (De Pontieu et al. 2014). We mainly use the slit-jaw images (SJIs) of 1400 {\AA} and
2796 {\AA} passbands, which have a cadence of 65 s, a pixel scale of 0.{\arcsec}3327, and a FOV of 167{\arcsec} $\times$
175{\arcsec}. The 1400 {\AA} channel samples emission of the Si {\sc iv} 1394 {\AA} and 1403 {\AA} lines formed in the
transition region (10$^{4.9}$ K) and the UV continuum emission from the lower chromosphere. The 2796 {\AA} channel is
dominated by the Mg {\sc ii} k 2796 {\AA} line emission formed in the upper chromosphere (10$^{4.0}$ K). For calculating
Doppler velocity of the jets, we also employ Si {\sc iv} 1402.77 {\AA} spectral line formed in the middle transition
region with a temperature of about 10$^{4.9}$ K. The nearby cold chromospheric S {\sc i} 1401.51 {\AA} line is assumed
to have a zero Doppler shift for the absolute wavelength calibration (Tian et al. 2018).

In addition, the observations from Atmospheric Imaging Assembly (AIA; Lemen et al. 2012) and Helioseismic and Magnetic
Imager (HMI; Schou et al. 2012) on board the \emph{SDO} are also used in the present work. The \emph{SDO}/AIA
successively observes the multilayered solar atmosphere in 10 passbands, including 7 extreme ultraviolet (EUV) channels
and 3 ultraviolet (UV) and visible channels. The \emph{SDO}/HMI provides full-disk line-of-sight (LOS) magnetograms,
intensitygrams, and photospheric vector magnetograms. Here we take the AIA full-disk 171 {\AA} images with a cadence
of 12 s and a spatial sampling of 0.{\arcsec}6 pixel$^{-1}$, HMI one-arcsecond resolution LOS magnetograms with a
cadence of 45 s, and the HMI vector data product called Space-weather HMI Active Region Patches (SHARP; Bobra et al.
2014). Furthermore, using the cross-correlation method, we carefully co-align data from the NVST, \emph{IRIS}, and
\emph{SDO} according to specific features that can be simultaneously detected in different channels.

In order to reconstruct 3D magnetic fields above the regions of interest, we utilize the ``weighted optimization"
method to perform NLFFF extrapolations (Wheatland et al. 2000; Wiegelmann 2004; Wiegelmann et al. 2012) based on the
HMI/SHARP photospheric vector magnetic fields observed at 01:00 UT on 2019 May 13 for AR 12741. The NLFFF extrapolation
is conducted within a box of $584\times 408\times128$ uniformly spaced grid points ($212\times148\times46$ Mm$^{3}$).
As the uncertainty of boundary condition will certainly give rise to unreliable extrapolation results, it is necessary
to assess the errors of the observed photospheric magnetic field employed here. The HMI/SHARP data product provides
the statistical error (standard deviation) for each vector field component, i.e., Bp\_err, Bt\_err, and Br\_err for
the phi component, theta component, and radial component of the vector magnetic field, respectively. At 01:00 UT on
2019 May 13, within the region of interest, the average values of phi component, theta component, and radial component
of the photospheric vector magnetic field are --774.7 G, 437.8 G, and --1210.3 G, and the average values of their
statistical errors are 48.4 G, --47.1 G, and 43.2 G. The uncertainties of the three components are about 6.2$\%$,
10.7$\%$, and 3.6$\%$, which are too small to significantly affect the extrapolation results (Zhu et al. 2017).
Furthermore, through the method developed by Liu et al. (2016), we calculate the squashing factor $Q$ of the
extrapolated fields, which provides important information about the magnetic connectivity (Titov et al. 2002).

\section{Results}
\subsection{Overview of the Sunspot Penumbral Filament Intrusions and Associated Jets}
The event of interest occurred within the main sunspot of AR 12741 when it was around the solar disk center on 2019
May 13. Figure 1(a) shows \emph{SDO}/HMI LOS magnetogram of this AR around 01:18 UT. The sunspot with negative magnetic
polarity is outlined by a green rectangle and enlarged to be shown in \emph{IRIS}/SJI 1400 {\AA} and \emph{SDO}/AIA 171
{\AA} images (panels (b)--(c)). It is shown that a light bridge divides the sunspot into two parts (see blue contours).
In 1400 {\AA} and 171 {\AA} channels, bright jet-like activities are observed to emanate from the west part of
this light bridge (also see the associated animation). Panels (d)--(g) display the light bridge and jets with a smaller
FOV (see the green rectangle in panel (c)) in NVST TiO and H$\alpha$ channels. The TiO image reveals that several
penumbral filaments deviate strongly from a radial configuration. They intrude into the west part of the light bridge
and form two groups of penumbral intrusions, whose tips penetrate into the sunspot umbrae from both sides of the light
bridge (see the green rectangle in panel (d)). In the H$\alpha$ observations, two groups of jets (jet1 and jet2) can
be clearly discerned (panels (e)--(g)), which manifest as dark absorption features in the images of H$\alpha$ wings
at $\pm$0.5 {\AA} and are less clear in the H$\alpha$ core image. Similar to the intruding penumbral filaments in the
aspect of morphology, jet1 and jet2 originate from different sides of the light bridge and then approach to each other
during their southwestward extensions to tens of Mm in the plane of the sky (POS). Moreover, brightenings are
intermittently detected in H$\alpha$ channels near the footpoints of the jets (see the red contours).

\subsection{Footpoint Brightenings and Kinematic Characteristics of the Jets}
\begin{figure*}
\centering
\includegraphics [width=0.75\textwidth]{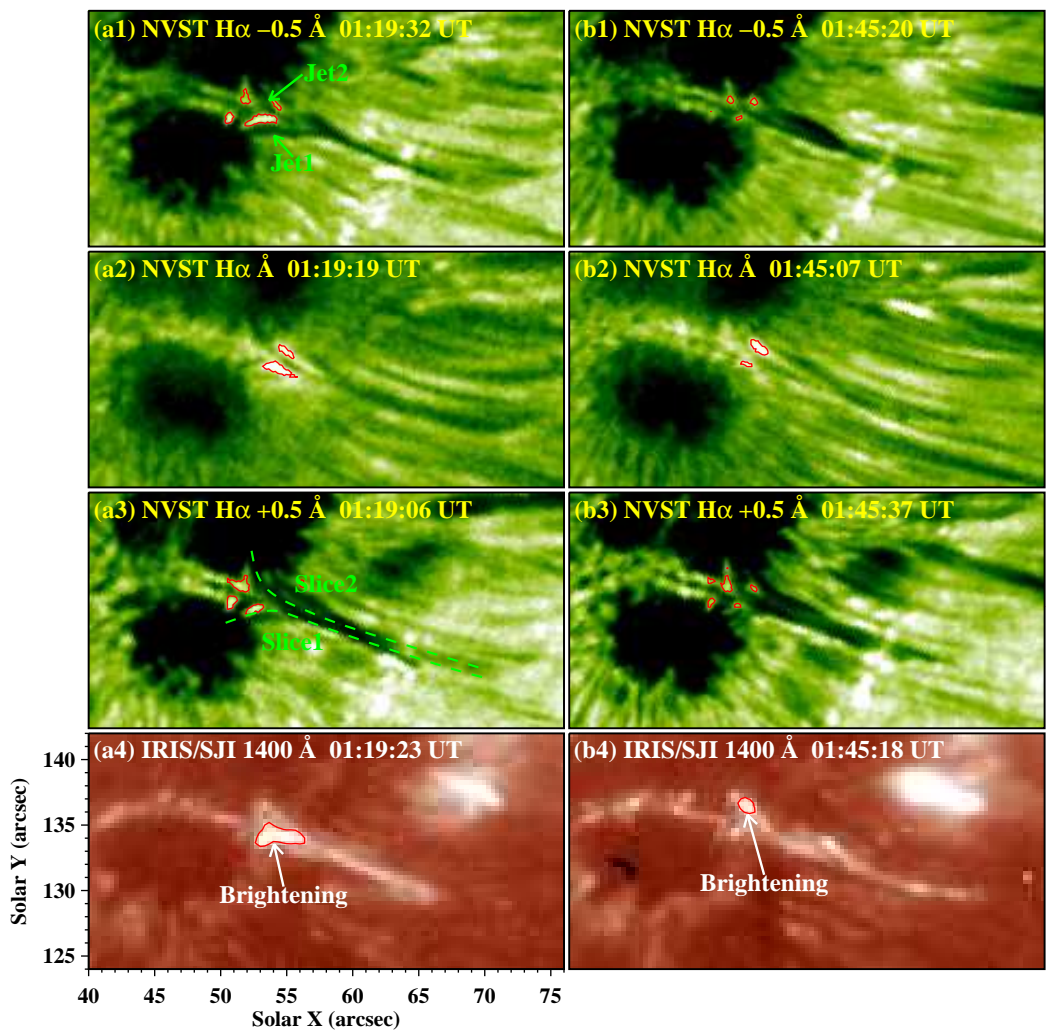}
\caption{Intermittent brightenings near the jet bases.
(a1)--(a4): NVST H$\alpha$ blue wing image ($\Delta\lambda=-0.5$ {\AA}), core image, red wing image
($\Delta\lambda=+0.5$ {\AA}), and \emph{IRIS} 1400 {\AA} image showing the jets and associated brightenings (see
the red contours) near the jet bases around 01:19 UT. The curves ``Slice1'' and ``Slice2'' in panel (a3)
approximate the projected trajectory of jet1 and jet2, respectively.
(b1)--(b4): Similar to (a1)--(a4), but for the time point of 01:45 UT.
An associated animation (figure2.mov) of H$\alpha$ blue wing, core, and red wing images, covering from 01:00 UT
to 02:28 UT, is available online.
}
\label{fig2}
\end{figure*}

As mentioned above, during the evolution of the jets, transient brightenings recurrently appeared around the jet base.
Figures 2(a1)--2(a3) show these brightenings around 01:19 UT in NVST H$\alpha$ blue wing, core, and red wing channels,
respectively (also see the associated animation). Signatures of these brightenings can also be identified from
simultaneous \emph{IRIS} 1400 {\AA} observations (see red contours in panel (a4)), indicating heating of local plasma
to about 10$^{4.9}$ K. Note that this temperature is applicable only when the assumption of coronal equilibrium
(CE) is valid (i.e., CHIANTI-based temperature) and may not be used for high density atmospheres such as Ellerman bombs
and UV bursts, where the formation temperature of \emph{IRIS} Si {\sc iv} lines could be 15000--20000 K (Rutten 2016;
Toriumi et al. 2017). These transient brightenings recurred around the base region of jets, followed by ejections of
the jets. The green dashed curves ``Slice1'' and ``Slice2'' in panel (a3) approximately denote projected trajectories
of jet1 and jet2 in the POS, respectively. Figures 2(b1)--2(b4) display the brightenings around 01:45 UT. Although
these brightenings had weaker emission enhancement than those around 01:19 UT, they were scanned by the
\emph{IRIS} spectral slit at that time, which enables us to have a comprehensive spectral analysis.

\begin{figure*}
\centering
\includegraphics [width=0.75\textwidth]{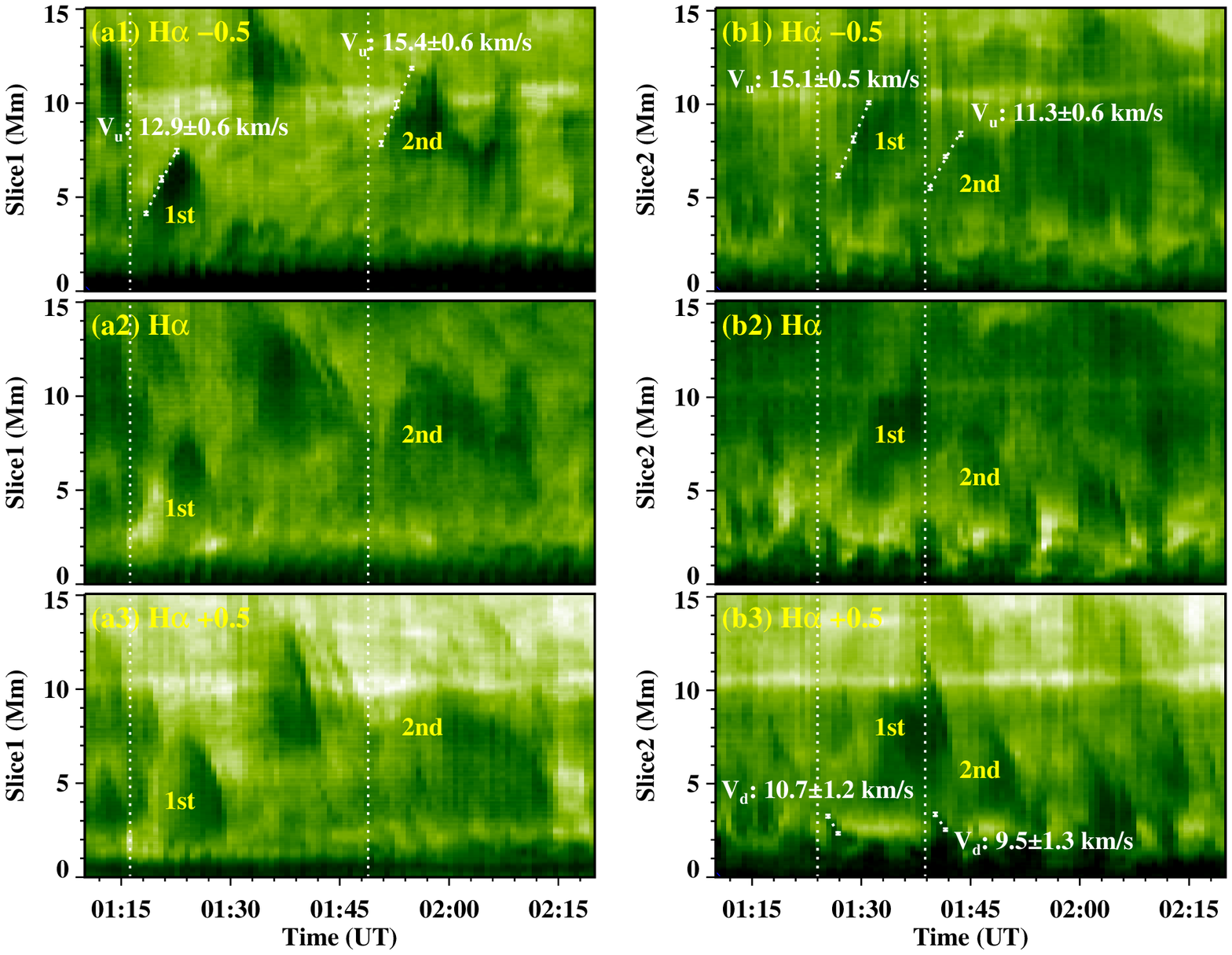}
\caption{Kinematic characteristics of the jets revealed by the NVST H$\alpha$ observations.
(a1)--(a3): Time-distance plots derived from the H$\alpha$ blue wing, line core, and red wing images along the
curve ``Slice1'' shown in Fig. 2(a3). The dotted vertical lines mark the initial moments of two rising phases.
The dotted oblique lines delineate linear fittings of the front of jet1 in two rising phases.
(b1)--(b3): Similar to (a1)--(a3), but showing jet2 along the curve ``Slice2".
}
\label{fig3}
\end{figure*}

To investigate the kinematic characteristics of jet1 and jet2, we derive time-distance plots from sequences of
H$\alpha$ images along the curves ``Slice1'' and ``Slice2'' in Fig. 2(a3) and then display them in Figs. 3(a1)--3(a3)
and Figs. 3(b1)--3(b3), respectively. It is shown that upward motions of the jets are clear in H$\alpha$ blue wing at
-0.5 {\AA} (panels (a1) and (b1)), and downward motions are distinct in H$\alpha$ red wing at +0.5 {\AA} (panels (a3)
and (b3)). Such kinematic characteristics of jets are also obviously revealed by the associated animation of Fig. 2.
Here we focus on the upward motions of jet1 and jet2, and two rising phases of each jet are investigated in detail,
onsets of which are marked by white dotted vertical lines. In the time-distance plots derived from H$\alpha$ blue wing
images, we track the jet front of each jet in the rising phase and record their y-axis coordinates at three time points
with set intervals. After repeating the track processing five times, we get the five groups of the jet front's
coordinates in each rising phase. Then we calculate the average values of these coordinates and plot three points at
the time-distance plot for each ascending phase. And the standard deviation is taken as the uncertainty. Finally, these
points are fitted through the linear fitting, which is shown as a dotted oblique line in Figs. 3(a1) and 3(b1). The
slope of the linear fitting line indicates the POS velocity of the jets. The projected ascending velocities of the front
motions in the POS are 12.9 $\pm$ 0.6 km s$^{-1}$ and 15.4 $\pm$ 0.6 km s$^{-1}$ for jet1, and are 15.1 $\pm$ 0.5
km s$^{-1}$ and 11.3 $\pm$ 0.6 km s$^{-1}$ for jet2. Note that these POS velocities are measured at the late stage of
the ascent, when the upward ejecting jet fronts have decelerated from their peak velocities at the initial ascent
stage. Furthermore, two downward flows are detected below the jet2 base with projected velocities of 10.7 $\pm$ 1.2
km s$^{-1}$ and 9.5 $\pm$ 1.3 km s$^{-1}$ (see panel (b3)). We can see that a base brightening accompanies some jet
fronts (such as the first one of jet1) while no emission enhancement is visible around the jet bases for the other
ones. This happens because the brightening regions around the jet bases sometimes are not crossed by ``Slice1'' and
``Slice2'', but which are visible in the associated animation.

\begin{figure*}
\centering
\includegraphics [width=0.75\textwidth]{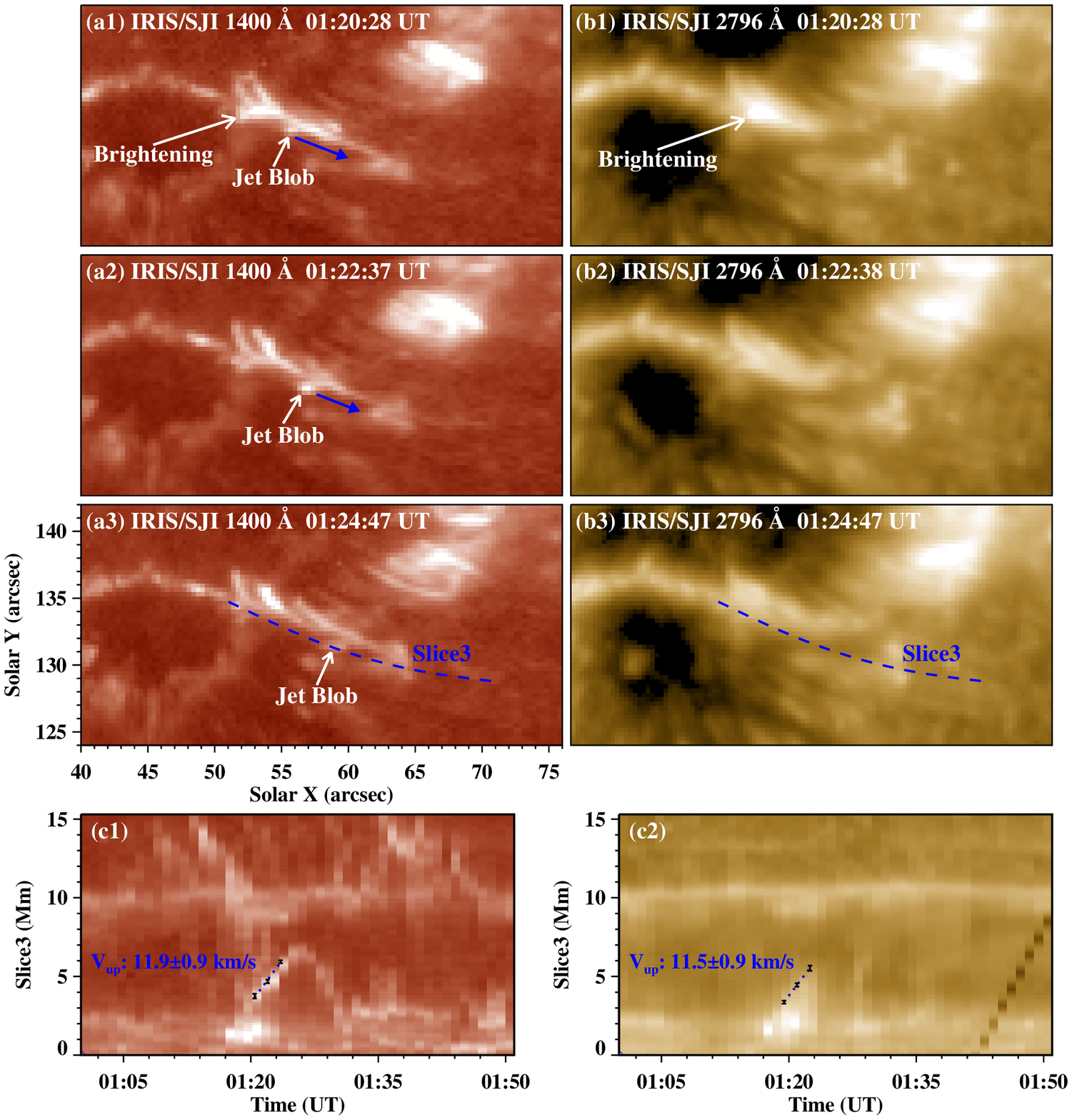}
\caption{Kinematic characteristics of the jets revealed by the \emph{IRIS} observations.
(a1)--(a3): Sequence of 1400 {\AA} images exhibiting the bright blob-like feature moving along the jet. The blue
solid arrow in panel (c) denotes the ejecting direction of the bright blob. The dashed curve ``Slice3''
approximates the projected trajectory of bright blob.
(b1)--(b3): Corresponding 2796 {\AA} images.
(c1)--(c2): Time-distance plots along the curve ``Slice3'' in 1400 {\AA} and 2796 {\AA} channels. The dotted
oblique lines delineate linear fittings of projected trajectory of the bright blob in the rising phase.
An animation (figure4.mov) of 1400 {\AA} and 2796 {\AA} channels, covering 00:40 UT to 01:51 UT, is available online.
}
\label{fig4}
\end{figure*}

Through checking simultaneous observations of the \emph{IRIS}, we find that the intermittent brightening around the
jet bases in H$\alpha$ channels can also be clearly detected in both of 1400 {\AA} and 2796 {\AA} images (see Figs.
4(a1) and 4(b1) and the associated animation). After the appearance of base brightening around 01:20 UT, a bright
blob-like feature was observed in 1400 {\AA} channel (see the white arrows in panels (a1)-(a3)). The bright blob
was ejected outward from the brightening site and moved along a POS trajectory similar to that of jet1 and jet2.
The blue dashed curves (``Slice3'') in panels (a3) and (b3) approximate the POS trajectory of the bright blob, along
which two time-distance plots are obtained from sequences of 1400 {\AA} and 2796 {\AA} images (see panels (c1) and (c2)).
It is shown that the POS motion of the bright blob in 1400 {\AA} time-distance plot matches well with a parabola
consisting of an upward phase and a falling phase, which appears to be in the nature of true mass motion. Taking the
same method used to estimate the POS velocities of the jet fronts in the H$\alpha$ time-distance plot, we find that
at the late stage of the ascent, the blob moved upward with a POS velocity of 11.9 $\pm$ 0.9 km s$^{-1}$, and finally
fell back. Note that the 1400 {\AA} emission of the blob during the upward phase is obviously stronger than that in the
falling one, indicating local heating possibly caused by shock fronts or compression related to the upwardly-ejecting
material during the ascent phase (Hou et al. 2017; Tian et al. 2018). However, the 2796 {\AA} time-distance plot of
panel (c2) reveals that the enhanced emission of the bright jet blob can only be identified during the period
corresponding to the ascent phase in panel (c1) with a POS velocity of 11.5 $\pm$ 0.9 km s$^{-1}$ and that the
trajectory is less distinct than that in 1400 {\AA} channel. Since the \emph{IRIS} 1400 {\AA} and 2796 {\AA} channels
mainly sample the emission from plasma with different temperatures, these observations imply that the
upwardly-ejecting materials are multi-thermal. And the materials heated to formation temperature of 1400 {\AA}
channel could be accelerated enough to reach the height of transition region and then fell back. But the materials
with lower temperature corresponding to 2796 {\AA} channel would fall after they reached much lower
heights, forming superimposed trajectories under a certain height.

\subsection{Spectral Properties of the Jets and the Footpoint Brightenings}
\begin{figure*}
\centering
\includegraphics [width=0.9\textwidth]{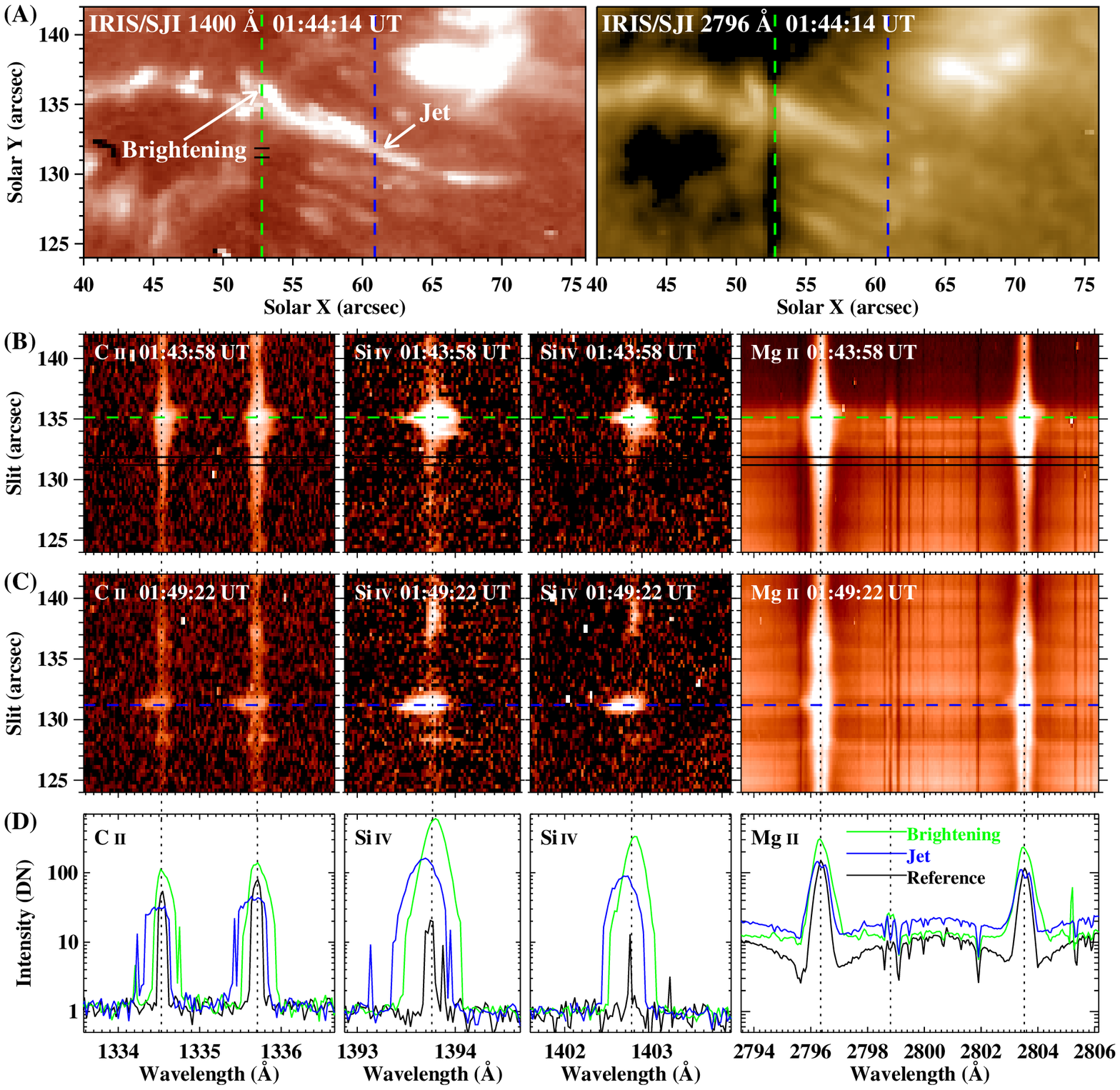}
\caption{\emph{IRIS} spectral observations of the jets and jet base brightening. (A): \emph{IRIS} 1400 {\AA} and
2796 {\AA} images. The green and blue vertical dashed lines mark the positions of spectrograph slit at 01:43:58 UT
and 01:49:22 UT, respectively. (B)--(C): Spectral Detector images taken through the spectrograph slit at the positions
shown in (A). (D): \emph{IRIS} spectral line profiles along the green horizontal dashed line in (B) at the jet base
brightening and the blue line in (C) at the jet indicated by the arrows in (A). The reference line profiles are
obtained by averaging the spectra within the section between the two black horizontal lines in (B) (also see the black
bars in (A)).
Here the intensities of profiles in Mg {\sc ii} window are all divided by 10.
}
\label{fig5}
\end{figure*}

From 01:42 UT to 01:51 UT, the \emph{IRIS} spectrograph slit scanned our focused region. The slit crossed the
brightening near the jet base around 01:43:58 UT (see green vertical dashed line in Fig. 5(A)). After that, the slit
scanned the jet from east to west, and we select a slit position at 01:49:22 UT (see blue vertical dashed line) to
investigate the spectral features of the jet. Panels (B) and (C) exhibit the detector images of C {\sc ii}
1334.53/1335.71 {\AA}, Si {\sc iv} 1393.76/1402.77 {\AA}, and Mg {\sc ii} h\&k 2796.35/2803.52 {\AA} windows taken
through the slit at the two positions shown in panel (A). We average the spectra within a relatively quiet region
around the jet base brightening (see the black bars in panel (A)) and take the resultant profiles as reference profiles
(see the black curves in panel (D)). Green and blue horizontal dashed lines in panels (B) and (C) indicate the locations
where the slit crosses the jet base brightening and jet. The profiles of UV emission lines of C {\sc ii}, Si {\sc iv},
and Mg {\sc ii} ions measured along these two horizontal lines are exemplified by green and blue curves in panel (D).
It is obvious that compared with the reference profiles, the profiles sampled at the location of the jet base
brightening (green curves) are dramatically enhanced and remarkably broadened at both wings of each emission line.
These features imply the existence of heating and bidirectional flows related to magnetic reconnections around the
jet base region (Peter et al. 2014; Tian et al. 2018). Another interesting feature for the jet base brightening is
that the Mg {\sc ii} 2798.809 {\AA} line, which is actually a blend of 2798.754 {\AA} and 2798.822 {\AA} lines,
changes from absorption to emission. Since the Mg {\sc ii} 2798.809 {\AA} line emission can serve as a diagnostic
for lower chromosphere heating (Pereira et al. 2015; Tian et al. 2015, 2016), this feature further demonstrates the
occurrence of magnetic reconnections around the jet base region. Furthermore, the profiles detected at the jet
(blue curves) present a distinct intensity enhancement of the blue wings, indicating flows towards the observer
along the LOS.

\begin{figure*}
\centering
\includegraphics [width=0.85\textwidth]{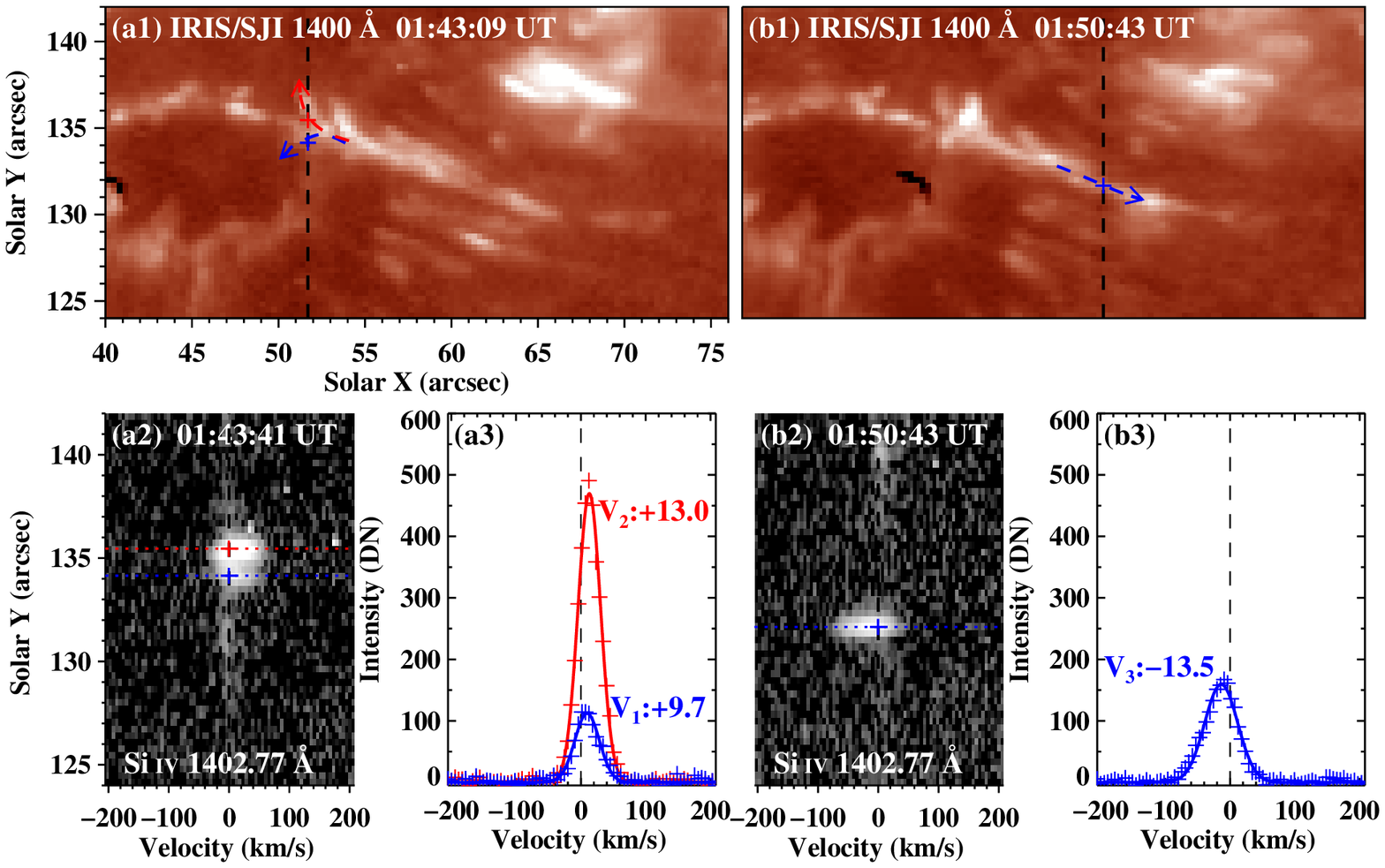}
\caption{Analysis of \emph{IRIS} Si {\sc iv} 1402.77 {\AA} line for the jets with two directions.
(a1)--(a3): Spectral analysis of the jets toward the solar surface. The black vertical line in (a1) marks the
location of spectrograph slit at 01:43:41 UT, and the arrows denote the directions of downward jet flows cross the
slit. Panel (a2) shows the Si {\sc iv} 1402.77 {\AA} line spectra along the slit. Panel (a3) exhibits the
Si {\sc iv} 1402.77 {\AA} line profiles (plus signs) and their single-Gaussian fitting profiles (solid curves) at
positions indicated by the blue and red plus signs (dotted lines) shown in (a2).
(b1)--(b3): Similar to (a1)--(a3), but for the jets ejecting upward.
}
\label{fig6}
\end{figure*}

To calculate the exact LOS velocity of the jets, we analyze the \emph{IRIS} Si {\sc iv} 1402.77 {\AA} line
profile. At 01:43:09 UT, the \emph{IRIS} spectrograph slit was located at the east of the jet base brightening and
crossed the bases of jet1 and jet2 observed in H$\alpha$ channels (see the blue and red plus symbols in Fig. 6(a1)).
At the two crossing positions, obvious redshift signals are exhibited in the Si {\sc iv} line spectra of panel (a2).
The red and blue plus signs in panel (a3) delineate the observed Si IV 1402.77 {\AA} profiles sampled along the red
and blue lines shown in panel (a2). Since the observed Si {\sc iv} profiles are close to Gaussian distribution, we
apply single-Gaussian fitting to approximate the line profile (see the red and blue solid curves in panel (a3)).
The Gaussian fit to the Si {\sc iv} line profile is obtained by computing a non-linear least-squares fit to a Gaussian
function $f(x)$ with three parameters: $f(x)=A_{0}e^{\frac{-z^{2}}{2}}$, where $z=\frac{x-A_{1}}{A_{2}}$. The three
parameters represent the height of the Gaussian ($A_{0}$), the center of the Gaussian ($A_{1}$), and the width of the
Gaussian ($A_{2}$), respectively. Then we get the Doppler velocity and its uncertainty according to the center of the
Gaussian ($A_{1}$) and its 1-sigma error estimate (Li et al. 2016; Zhang et al. 2018a). The Doppler velocities at the
blue and red plus positions are 9.7 $\pm$ 0.5 km s$^{-1}$ and 13.0 $\pm$ 0.9 km s$^{-1}$, indicating downward flows at
the east of the base brightening of jet1 and jet2 (see blue and red dashed arrows in panel (a1)). At 01:50:43 UT, the
scanning slit crossed the jet at a higher height, where jet1 and jet2 had merged into one jet flow (panel (b1)). The
spectroscopic analysis reveals that there is a significant blueshift signal of --13.5 $\pm$ 0.4 km s$^{-1}$ at the
crossing position (panels (b2)--(b3)), implying an upward flow (see the blue dashed arrow in panel (b1)).

\subsection{3D Magnetic Fields of the Penumbral Filaments Revealed by NLFFF Extrapolation}
\begin{figure*}
\centering
\includegraphics [width=0.8\textwidth]{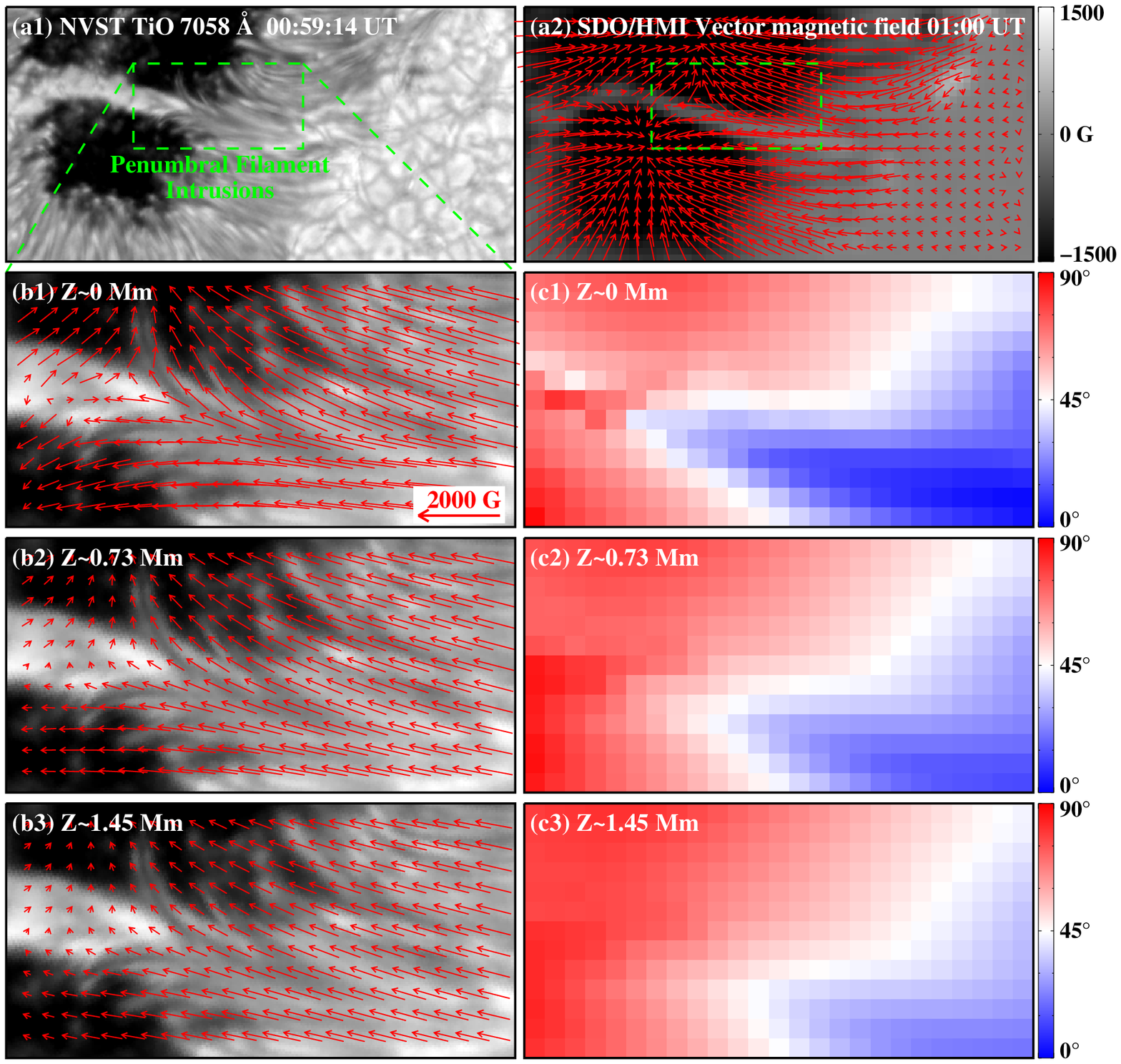}
%[bb=20 210 580 630,clip,angle=0,width=0.89\textwidth]{figure6.eps}
\caption{Magnetic fields of the intruding penumbral filaments and sunspot light bridge.
(a1)--(a2): NVST TiO image and SDO/HMI vector magnetogram showing the sunspot penumbral filaments intruding into
umbrae at both sides of the light bridge and the corresponding photospheric magnetic fields.
(b1)--(b3): Enlarged TiO images superimposed by the horizontal magnetic fields (red arrows) calculated by NLFFF
extrapolation at the height of z$\simeq$0 Mm, z$\simeq$0.73 Mm, and z$\simeq$1.45 Mm, respectively.
(c1)--(c3): Corresponding maps of angle between the vector magnetic field and the horizontal plane.
}
\label{fig7}
\end{figure*}

The kinematic characteristics of the jets ought to be accounted for magnetic fields of the intruding penumbral
filaments and light bridge, which will be further analyzed in Figs. 7 and 8. The NVST TiO image at 00:59:14 UT in
Fig. 7(a1) shows that superposed on the west part of the light bridge, several penumbral filament intrusions penetrate
into the umbrae from both sides. Each penumbral filament manifests as a dark core flanked by lateral brightenings.
The corresponding photospheric vertical magnetic field is plotted as the background in panel (a2), and the red arrows
indicate the photospheric horizontal magnetic field around 01:00 UT. In the enlarged TiO image of panel (b1), red
arrows clearly depict strong photospheric horizontal fields ($\simeq$1000 G) along the penumbral filaments. At the
east part of the light bridge, relatively weaker and almost horizontal fields with an opposite direction are also
detected. Furthermore, we compute the angle $\theta$ between the vector magnetic field and the horizontal plane as
follows:
\begin{equation}
\theta=\arctan\frac{|B_{z}|}{\sqrt{B_{x}^{2}+B_{y}^{2}}}
\label{eq1}
\end{equation}
The $\theta$ map shown in panel (c1) reveals that at the photospheric layer (Z$\simeq$0 Mm), the sunspot umbrae have
magnetic fields close to vertical ($\theta\simeq$90{\degr}), and the fields of sunspot penumbrae are more inclined
($\theta\simeq$45{\degr}). The magnetic fields of penumbral filaments intruding into the light bridge are nearly
horizontal ($\theta\simeq$0{\degr}). To reconstruct 3D magnetic fields of AR 12741, we then perform NLFFF modeling
based on the observed photospheric vector fields at 01:00 UT. Panels (b2)--(c2) and (b3)--(c3) display the
extrapolated horizontal fields and $\theta$ maps at the height of Z$\simeq$0.73 Mm and Z$\simeq$1.45 Mm, respectively.
Note that with the increase of altitude, the range of penumbral filament horizontal fields (deep blue region in the
$\theta$ maps) keeps shrinking southwestward and gradually retreats from the light bridge. It indicates that, above
the light bridge region, the higher the height is, the larger the $\theta$ value of the intruding filament fields is.
As for the southwest parts of the filaments, which are out of the light bridge region, they are dominated by
inclined magnetic fields with smaller $\theta$ than surrounding penumbrae even at the height of Z$\simeq$1.45 Mm.

\begin{figure}
\centering
\includegraphics [width=0.49\textwidth]{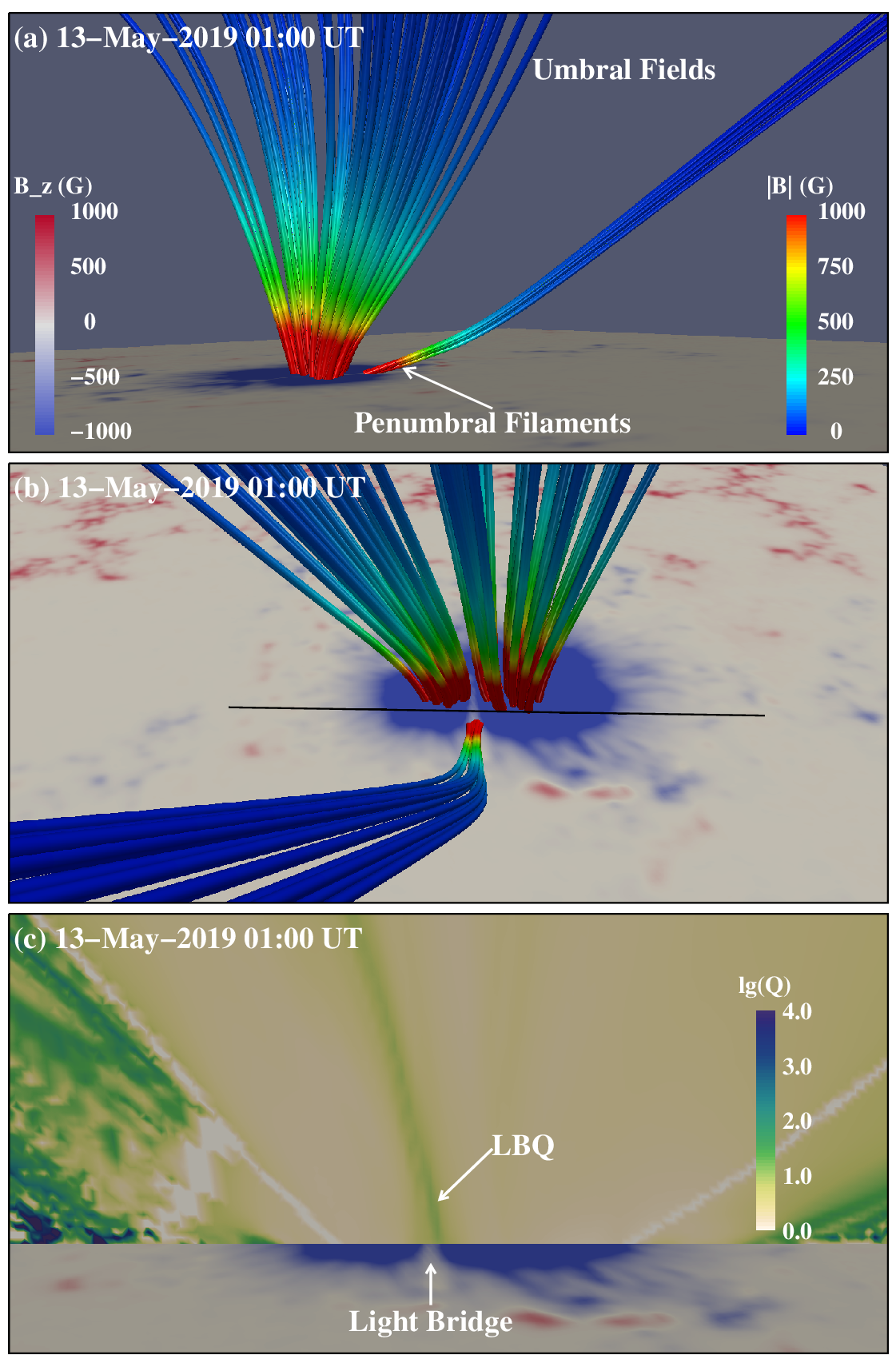}
\caption{3D magnetic topology of the sunspot penumbral filaments intruding into the light bridge revealed by NLFFF
extrapolation at 01:00 UT on 2019 May 13. Panels (a) and (b) show the intruding filaments and sunspot umbral fields
from a side view and a top view, respectively. Panel (c) displays logarithmic $Q$ distribution in the vertical plane
based on the black cut denoted in panel (b), which distinctly depicts the high-$Q$ region above the light bridge (LBQ).
}
\label{fig8}
\end{figure}

For visualizations of the 3D magnetic topologies of our focused structures, we select a region from the NLFFF
extrapolation and display it in Fig. 8. Moreover, we calculate the distributions of strength $|B|$ and squashing
factor $Q$ of the reconstructed fields. Figures 8(a) and 8(b) show the sunspot umbral fields and intruding penumbral
filaments from the side view and top view, respectively. The plotted field lines have been colored depending on the
local magnetic field strength ($|B|$). It is clear that there are highly inclined fields (stronger than 1000 G)
intruding into the light bridge. These magnetic fields should correspond to the observed intruding penumbral filaments.
In the $Q$ map of vertical plane above a cut crossing the light bridge (see black line in panel (b)), a high-$Q$ region
above the light bridge (LBQ) is visible (panel (c)), implying the existence of magnetic canopy.

\section{Summary and Discussion}
Based on high-quality observations from the NVST, \emph{SDO}, and \emph{IRIS}, we investigate an interesting event where
several penumbral filaments intruded into a sunspot light bridge. On 2019 May 13, a light bridge completely separated the
umbra of the main sunspot in AR 12741. The NVST TiO images show that superposed on the west part of the light bridge,
several penumbral filaments intrude into the sunspot umbrae from both sides of the light bridge. Meanwhile, in \emph{IRIS}
1400 {\AA} and \emph{SDO}/AIA 171 {\AA} channels, bright jet-like activities were observed to emanate from the west part of
this light bridge. Similarly, the NVST H$\alpha$ observations, especially at two wings of $\pm$0.5 {\AA}, reveal that two
groups of jets (jet1 and jet2) originated from different sides of the light bridge and then approached to each other when
moving southwestward in the POS, which shared the same projected morphology with the intruding filaments. Additionally,
intermittent brightenings and downward flows were detected around the jet bases. The \emph{IRIS} spectral observations
also provide convincing evidence for the presence of magnetic reconnection related heating and bidirectional flows
near the jet base region. Moreover, the \emph{IRIS} SJIs of 1400 {\AA} and spectral observations of 1402.77 {\AA} line
reveal that at the formation height of 1402.77 {\AA} line, the jets ejected southwestward in the POS had a POS velocity of
11.9 $\pm$ 0.9 km s$^{-1}$ around 01:20 UT, and the LOS velocity was estimated as 13.5 $\pm$ 0.4 km s$^{-1}$ at 01:50 UT.
In the aspect of magnetic topology, the observed photospheric magnetic fields and extrapolated 3D fields reveal the
existence of strong and highly inclined magnetic fields along the intruding penumbral filaments.

The jets reported in the present work can be observed in H$\alpha$, 1400 {\AA}, and 171 {\AA} channels, indicating that
the jets consist of multi-thermal components. Moreover, the jets are accompanied by intermittent brightenings around the
bases, which are clearly visible in the 1400 {\AA} and 2796 {\AA} images. It means that the local material at the jet base
region is heated to at least 10$^{4.0}$ K. In addition, the \emph{IRIS} spectral profiles of the jet base
brightening are significantly enhanced and broadened at both wings of each emission line. These features are very similar
to the recent \emph{IRIS} observation of brightening events and are interpreted as the local heating of plasma via magnetic
reconnection and the bidirectional outflows from the reconnection region (Peter et al. 2014; Toriumi et al. 2015a, 2017;
Tian et al. 2016, 2018; Huang et al. 2018; Young et al. 2018; Chen et al. 2019). Moreover, the Mg {\sc ii} 2798.809 {\AA}
line of the jet base brightening changes from absorption to emission. In solar spectra, Mg {\sc ii} 2798.809 {\AA} line
is seen mostly in absorption, but can become emission lines when strong heating takes place in the lower chromosphere
(Pereira et al. 2015; Tian et al. 2015, 2016; Toriumi et al. 2017). Therefore, it is highly possible that the profiles
shown in Fig. 5 indicate the energy release caused by magnetic reconnection at the jet base region. It is worth
mentioning that reconnection is not the only mechanism that can lead to the broadening of the spectra line. Currents,
for instance, can also produce dissipation that can yield similar spectral shapes. However, the intensities of the
spectra lines analyzed in this work are so strong that they cannot be mainly caused by the currents. And the
strong enhancements to both blue and red wings may not be explained merely by the Joule dissipation. As a result,
we propose that the studied southwestward jets in the POS should probably be outflows in one direction caused by the
reconnection occurring around the jet bases. This interpretation is further supported by the downward flows displayed
in Fig. 3(b3), which may represent the outflow toward another direction.

In Fig. 3, we estimate the projected ascending velocities of the jets through the time-distance plot of H$\alpha$
observations and obtain the values of 12.9 $\pm$ 0.6 km s$^{-1}$, 15.4 $\pm$ 0.6 km s$^{-1}$, 15.1 $\pm$ 0.5 km s$^{-1}$,
and 11.3 $\pm$ 0.6 km s$^{-1}$ at different time points. But for a vector velocity of the jets, we need to get the LOS
component of true velocity, which can be measured by the \emph{IRIS} spectral observations. Around 01:20 UT, jets were
observed accompanied by distinct footpoint brightenings in both H$\alpha$ and 1400 {\AA} channels. Figure 4 reveals that
a bright jet blob feature detected in \emph{IRIS} 1400 {\AA} channel moved southwestward with a POS velocity of about
11.9 km s$^{-1}$, which is comparable to the value of 12.9 $\pm$ 0.6 km s$^{-1}$ that calculated based on the NVST
H$\alpha$ observations. However, due to the 320-step raster mode of the IRIS spectrograph slit, the LOS velocity of the
jet can only be obtained when the scanning slit crossed the jet after 01:45 UT, whereas the POS motion of the
jet was not distinct in the 1400 {\AA} slit-jaw images but clear in H-alpha blue wing observations (2nd jet in
Fig. 3(a1)). Thus for the jet around 01:50 UT shown in Fig. 6, which had a LOS velocity of $\sim$13.5 km s$^{-1}$
estimated through the Doppler shift of Si {\sc iv} 1402.77 {\AA} line, we adopt the POS velocity ($\sim$15.4 km s$^{-1}$)
derived from the H-alpha observations as an approximation to that in the 1400 {\AA} observations. Therefore, the true
velocity of the jet will be estimated to be about 20.5 km s$^{-1}$. Considering that the velocity is measured during the
late stage of the ascent near the transition region, where the upward jets have been significantly decelerated by the
solar effective gravity or resistance from the upper atmosphere, the estimated value should be the lower limit value of
the initial speed of the upward jets when they are launched from a lower height, e.g., upper photosphere or lower
chromosphere. That is, the peak velocity of the jet at the initial ascent stage should be much larger than
20.5 km s$^{-1}$.

If we assume a proton number density ($n$) of 10$^{15}$ cm$^{-3}$ and a magnetic field strength ($B$) of 500 Gauss near
the lower chromosphere above the sunspot light bridge, the Alfv\'{e}n speed ($V_{A}$) could be estimated to be $\sim$ 34.5
km s$^{-1}$ according to the formula $V_{A}=B / \sqrt{\mu m_{p} n}$. Since the estimated Alfv\'{e}n speed is not much
larger than the lower limit value (20.5 km s$^{-1}$) of the initial speed of the observed jets, it is reasonable to suggest
that these jets are produced by magnetic reconnection near the lower solar atmosphere above the light bridge.
Additionally, at a temperature of 10000 K of the lower solar atmosphere, the local sound speed ($C_{s}\simeq
152T^{1/2}$ m s$^{-1}$) could be estimated as $\sim$ 15.2 km s$^{-1}$, which is smaller than the speed of the observed
jets. It further supports that intensity enhancement of the jet during its upward phase shown in Fig. 4 is caused by
reconnection-related shock heating. Moreover, the intermittent and aperiodic occurrences of these jets also exclude the
possible role of the p-mode wave on driving the jets, which is often suggested to account for the oscillating light walls
above light bridges with a dominant period of several minutes (Yang et al. 2015; Zhang et al. 2017; Hou et al. 2017;
Tian et al. 2018). One may see from the time-distance plots in Fig. 3 that the jets have a parabolic trajectory
with a typical lifetime of 10--20 minutes. They are very similar to the surges reported by Toriumi et al. (2015a) and
Robustini et al. (2016), which are interpreted as results of repeated magnetic reconnection driven by magneto-convective
evolution.

\begin{figure*}
\centering
\includegraphics %[width=0.89\textwidth]{figure5.eps}
[bb=10 10 590 380,clip,angle=0,width=0.9\textwidth]{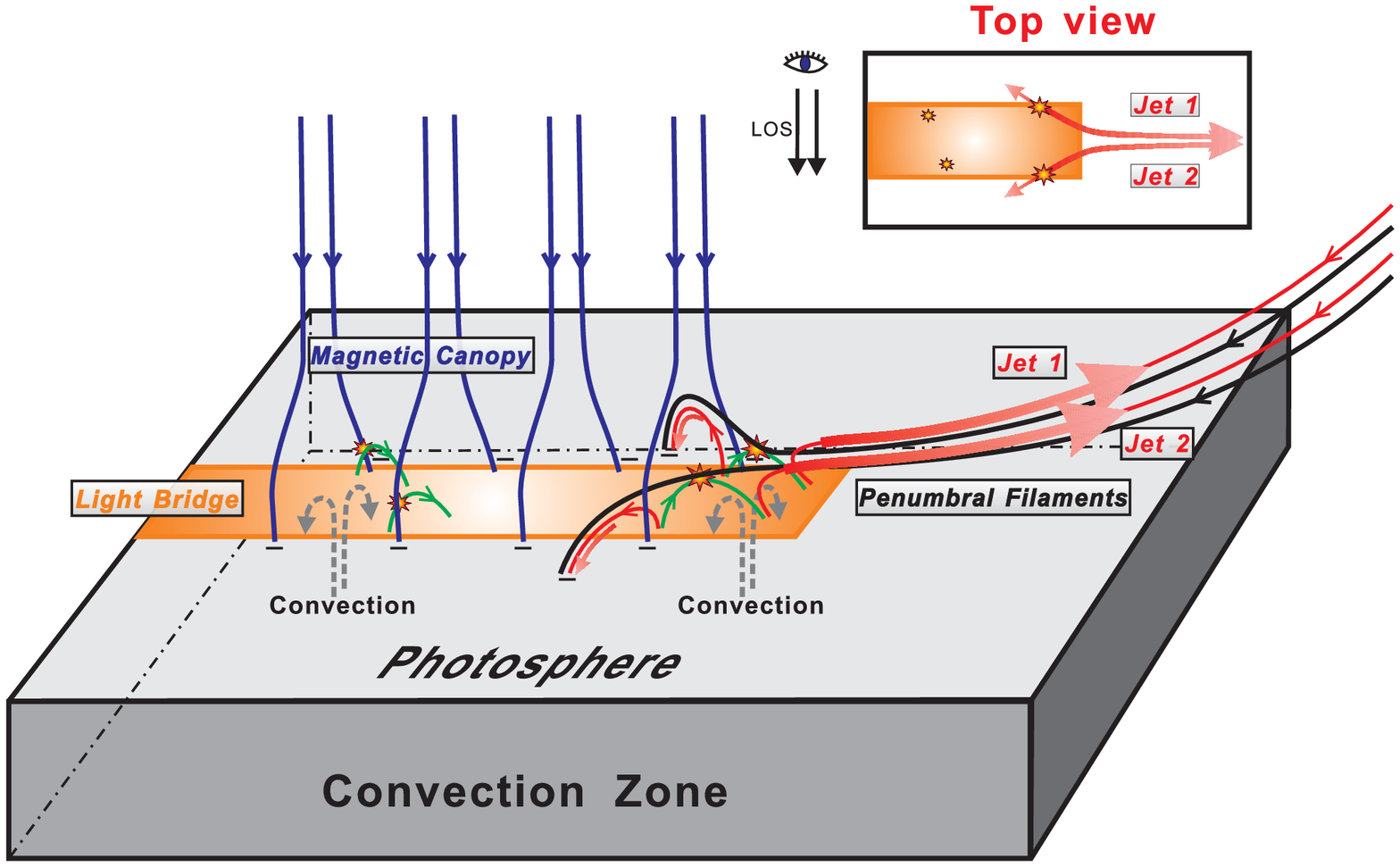}
\caption{Sketch illustrating formations of the jets caused by the sunspot penumbral filament intrusions into the sunspot
light bridge. The orange rectangle approximates mapping of the light bridge on the photosphere (indicated by the upper
surface of a grey cube). The green curves represent emerging magnetic fields within the light bridge. Umbral magnetic
fields surrounding the light bridge are denoted by the navy blue curves, which present as a magnetic canopy. The black
curves delineate magnetic fields of the penumbral filaments intruding into the umbrae on two sides of the light bridge.
The star symbols mark the sites of magnetic reconnection. The red curves denote newly formed magnetic fields due to the
reconnection occurring between the emerging fields within the light bridge and the highly inclined fields of the
intruding filaments, along which downward flows and two groups of upward jets (jet1 and jet2) are observed.
}
\label{fig9}
\end{figure*}

As modeled by Nakamura et al. (2012), the jets generated by magnetic reconnection tend to move along stronger magnetic
fields for that weaker fields have higher gas pressure in the initial equilibrium and hence result in the gas pressure
gradient along the strong fields, accelerating the field-aligned plasma flow. As a result, it is reasonable that within
sunspots, the surge-like activities reported earlier are all aligned with stronger and relatively vertical background
sunspot (umbral or penumbral) fields, because they are driven by the magnetic reconnection occurring either between
emerging magnetic fields within the light bridge and background sunspot (umbral) fields (Shimizu et al. 2009; Robustini
et al. 2016; Hou et al. 2017; Tian et al. 2018) or between the horizontal penumbral filament fields and background sunspot
(penumbral) fields (Katsukawa et al. 2007a; Jur{\v{c}}{\'a}k \& Katsukawa 2008; Magara 2010). But in the present work,
the studied jets have the same projected morphology as the highly inclined intruding penumbral filaments, whose fields
are weaker than the background sunspot fields but stronger than the emerging fields within the light bridge. So we
propose that these jets could be driven by the magnetic reconnection between emerging fields within the light bridge and
the intruding filament horizontal fields, and then were ejected outward along the stronger filament fields. As a result,
the answers to the two questions raised in the Introduction would be that if penumbral filaments intrude into the light
bridge, magnetic reconnection could occur between the emerging fields within light bridge and penumbral filament fields,
and jets along the stronger filament fields will be produced.

Penumbral filaments and light bridges are dominant radial structures in the sunspot penumbra and remarkable bright lanes
dividing the sunspot umbrae, respectively. The penumbral filaments are observed to have highly inclined or horizontal
magnetic fields (Louis et al. 2014b). And the curved filamentary structures are present in some light bridges (Rimmele 2008).
By analyzing high-resolution observations from the Swedish 1m Solar Telescope, \emph{SDO}, and \emph{Hinode}, Robustini
et al. (2016) reported fan-shaped jets above the light bridge of a sunspot driven by reconnection and noticed that this
light bridge exhibits a filamentary structure somewhat similar to the penumbral filament intrusions. However, due to the
lack of high-resolution TiO observations, the emission features and morphologies of the filamentary structures within the
light bridge and their relations to the fan-shaped jets cannot be investigated in detail. In the present work, with the
help of high-resolution NVST TiO observations, we clearly exhibit that the filamentary structures within the light bridge
here have typical penumbral filament features of dark cores flanked by lateral brightenings (Scharmer et al. 2002). These
filaments emanate from the sunspot penumbra, intrude into the light bridge, and then form two groups of penumbral
intrusions, whose tips penetrate into the sunspot umbrae from both sides of the light bridge (see Figs. 1(d) and 7(a1)).
The photospheric vector magnetic fields observed by \emph{SDO}/HMI reveal that the intruding penumbral filaments have
strong horizontal magnetic fields ($\simeq$1000 G) at the photospheric layer (see Figs. 7(b1) and 7(c1)). Moreover, the
NLFFF extrapolation results also provide solid evidences for the existence of highly inclined magnetic fields of the
penumbral filaments at higher positions above the photosphere (see Figs. 7(b2)-7(c3) and 8). It can be estimated that
near the transition region, i.e., the formation height of 1402.77 {\AA} line, the angle $\theta$ between the penumbral
filament field and the horizontal plane is 30{\degr}--50{\degr}. Based on the position of the jets on the solar disk and
the POS and LOS velocities of jets measured through the \emph{IRIS} 1402.77 {\AA} spectral imaging observations, we can
calculate the angle $\varphi$ between the jets and the local horizontal plane by applying the method of transforming
observed magnetic field vector in the image plane into the heliographic plane (Hagyard 1987; Gary \& Hagyard 1990). The
angle $\varphi$ between the moving direction of the jets and the local horizontal plane at the formation height of
1402.77 {\AA} line is estimated to be $\sim$40{\degr}, which is consistent well with the magnetic topology of the
penumbral filaments. These results further reinforce our conclusion that the jets emanating from the light bridge could
be moving along the fields of the penumbral filaments.

To illustrate the magnetic fields of intruding penumbral filaments and light bridge as well as the formation process of
the jets detected by the NVST, \emph{SDO}, and \emph{IRIS}, we draw a cartoon map and show it in Fig. 9. Within the
light bridge (see the orange rectangle), low-lying magnetic structures (see the green curves) continuously emerge
through long-lasting convection upflows from the solar interior to the solar atmosphere (Lites et al. 1991; Rueedi et
al. 1995; Leka 1997; Jur{\v c}{\'a}k et al. 2006; Rouppe van der Voort et al. 2010; Toriumi et al. 2015a, b; Zhang et
al. 2018b). They are then restricted in the lower solar atmosphere by the above strong umbral fields (see the blue
curves), which form a magnetic canopy covering the light bridge (Jur{\v c}{\'a}k et al. 2006). When part of these
emerging fields meet the overlying magnetic canopy field with opposite polarity at one side, current sheets are formed,
leading to magnetic reconnection (see the stars at the east part of the light bridge) in the lower atmosphere (Asai et
al. 2001; Shimizu et al. 2009; Hou et al. 2017; Tian et al. 2018). The resultant brightenings can be detected in
H$\alpha$, 1400 {\AA}, and 2796 {\AA} channels in this event (see the animations attached to Figs. 1, 2, and 4). At
the west part of the light bridge, the nearly horizontal fields of penumbral filaments (see the black curves) protrude
into the light bridge. When the horizontal components of the emerging fields within the light bridge and filament fields
are in opposite directions, current sheets will be formed. Then the vigorous convective upflows within the light bridge
will push field lines with opposite directions together at the current sheets, giving rise to repeated magnetic
reconnections between the horizontal filament fields and the emerging fields within the light bridge. The newly-formed
filament fields through reconnection are marked by the red curves and their footpoints within the sunspot keep moving
from the umbrae into the light bridge, which is also confirmed by the evolution of the intruding penumbral filaments
observed by the NVST. The repeated reconnections cause the observed intermittent brightenings while the reconnection
outflows with opposite directions are observed as upward jet1 and jet2 along the filament fields and downward flows
below the reconnection sites. From the top view, these outflows form a Y-shaped structure lying on the solar surface.

Noting that due to the lack of high-resolution observations of vector magnetic fields, such as the \emph{Hinode} data,
here we cannot find the direct evidence of the emerging magnetic fields within the light bridge. However, the emergence
of magnetic flux within the light bridge has been widely demonstrated in recent observational and numerical works
(Louis et al. 2015; Toriumi et al. 2015a, b; Yuan \& Walsh 2016; Tian et al. 2018; Dur{\'a}n et al. 2020). Furthermore,
it might be wondered that why within the same light bridge, the magnetic reconnection could trigger jets in some locations
(the west part of the light bridge in the reported event) but can not in some other places (the east part). We speculate
that it could be determined by the overlying magnetic topology and the occurrence height of the reconnection (deciding
the surrounding plasma density), which is planned to be investigated in detail in our further work.

Similar to our observations, Yang et al. (2019a) reported that at the edge of a light bridge, light bridge
filaments intruded into the sunspot umbrae. But the difference is that, around the tips of the intruding filaments,
photospheric vortices and vertically launched jets were detected by Yang et al. (2019a). These authors considered that
the strong transverse velocity shear between the light bridge and the umbrae triggered the Kelvin-Helmholtz instability
(KHI) resulting in the vortex structures, which then launched the jet in the vertical direction through magnetic
reconnection or waves. To our knowledge, the velocity shear larger than (it would be twice under certain assumptions)
local Alfv\'{e}n speed (here is ~10 km s$^{-1}$ in the photosphere) is needed to make KHI take place (Chandrasekhar 1961;
Li et al. 2019). And the KHI-induced vortices would form randomly at the boundary of flows (here is the edge of the light
bridge). Considering the actual observations in both events reported in Yang et al. (2019) and our study, we speculate that
the vortices could be caused by the interaction between the large-scale, long-lasting horizontal flows and some small-scale
local convection cells (Toriumi et al. 2015a) at some sites of the bridge during its developing phase, when the local
convections are relatively stronger than that in the mature (or decaying) bridges. The diverging flows caused by the
small-scale local convection would block the large-scale horizontal flows along the bridge filaments at some sites and
squeeze them (also the magnetic  fields) outwards, which then simultaneously intrude into the umbrae and form the vortex
structures at both sides of the bridge. Subsequently, the horizontal intruding filament fields would reconnect with the
vertical umbral fields and produce jets along the stronger vertical umbral fields. But for the mature light bridge
reported in our study, we suggest that the small-scale local convections are not strong enough to keep pushing the
filaments. Thus the intruding filaments observed here would be much less active than that in Yang et al. (2019) and thus
not reconnect with the umbral fields. On the other hand, random flux emergence caused by the weak local convection within
the light bridge would interact with the above horizontal filament fields, producing jets ejecting outward along the
stronger filaments fields. During the observation period of our event, we cannot find vertical jets in the light bridge.
But we believe that during the developing phase of the light bridge, when the small-scale local convections are strong,
the reconnection with the umbral field could be frequent, and thus vertical jets would be observed.

The magnetic fields of sunspot penumbra have been extensively investigated through observations of spectral lines
forming in the photosphere and numerical simulations, which indicate horizontal magnetic fields along dark penumbral
filaments and more vertical fields in surrounding regions of penumbra (Thomas et al. 2002; Langhans et al. 2005;
Scharmer et al. 2008; Rempel et al. 2009). Scharmer et al. (2002) presented the first observations of penumbral
filaments with clear dark cores. Spruit \& Scharmer (2006) introduced the concept of field-free convection gaps to
explain the penumbral filaments, and a strong horizontal magnetic field was predicted to exist along the penumbral
filament. Analysis in the present work sheds light on magnetic fields of the sunspot penumbral filament and light
bridge, as well as the interactions between them. It is revealed that the fields of the penumbral filaments
intruding into the sunspot light bridge and umbrae are indeed nearly horizontal at the photosphere and highly
inclined at higher layers, which are depicted by the jets along these filaments observed in NVST H$\alpha$ images.
The interpretation of formation of these jets mentioned above indicates that magnetic reconnection could occur
between the penumbral filament fields and emerging fields within the light bridge and results in jets along the
stronger filament fields. These results further complement the study of magnetic reconnection and dynamic
activities within the sunspot.

\begin{acknowledgements}
The authors appreciate the anonymous referee for the valuable suggestions. Y.H. thanks Prof. Hui Tian, Dr. Xiaoshuai
Zhu, Dr. Yongliang Song, and Dr. Shuo Yang for helpful discussions. The data are used courtesy of NVST, \emph{SDO}, 
and \emph{IRIS} science teams. \emph{SDO} is a mission of NASA's Living With a Star Program. \emph{IRIS} is a NASA 
small explorer mission developed and operated by LMSAL with mission operations executed at NASA Ames Research center 
and major contributions to downlink communications funded by ESA and the Norwegian Space Centre. The authors are 
supported by the National Key R\&D Program of China (2019YFA0405000), the National Natural Science Foundation of 
China (11533008, 11903050, 11790304, 11773039, 11873059, 11673035, 11673034, and 11790300), the Strategic Priority 
Research Program of the Chinese Academy of Sciences (XDB41000000), the NAOC Nebula Talents Program, the Youth 
Innovation Promotion Association of CAS, Young Elite Scientists Sponsorship Program by CAST (2018QNRC001), and Key 
Programs of the Chinese Academy of Sciences (QYZDJ-SSW-SLH050).
\end{acknowledgements}

% WARNING
%-------------------------------------------------------------------
% Please note that we have included the references to the file aa.dem in
% order to compile it, but we ask you to:
%
% - use BibTeX with the regular commands:
%   \bibliographystyle{aa} % style aa.bst
%   \bibliography{Yourfile} % your references Yourfile.bib
%
% - join the .bib files when you upload your source files
%-------------------------------------------------------------------

\clearpage

\end{document}